\documentclass[10pt,conference]{IEEEtran}

\usepackage{lgrind}
\usepackage{epsfig}
\usepackage{amsmath}    
\usepackage{graphicx}
\usepackage{amssymb}
\usepackage{fancyhdr}
\pdfoutput=1

\begin{document}

\title{Receiver design to harness quantum illumination advantage}

\author{
\authorblockN{Saikat Guha}
\authorblockA{{\em Disruptive Information Processing Technologies} \\
BBN Technologies \\
10 Moulton Street, Cambridge, MA 02138 \\
sguha@bbn.com}
}

\maketitle

\begin{abstract}
An optical transmitter that uses entangled light generated by spontaneous parametric downconversion (SPDC), in conjunction with an optimal quantum-optical receiver (whose implementation is not yet known) is in principle capable of obtaining up to a 6 dB gain in the error-probability exponent over the optimum-reception un-entangled coherent-state lidar to detect the presence of a far-away target subject to entanglement-breaking loss and noise in the free-space link \cite{lloyd2008, tan2008}. We present an explicit design of a structured quantum-illumination receiver, which in conjunction with the SPDC transmitter is shown to achieve up to a 3 dB error-exponent advantage over the classical sensor. Apart from being fairly feasible for a proof-of-principle demonstration, this is to our knowledge the first structured design of a quantum-optical sensor for target detection that outperforms the comparable best classical lidar sensor appreciably in a low-brightness, lossy and noisy operating regime. 
\end{abstract}

\section{Introduction}

An optical transmitter is employed to interrogate a distant region engulfed in bright thermal light, suspected of containing a weakly reflecting target. The return light is processed by a receiver to decide whether or not the target is present. Recently \cite{lloyd2008}, Lloyd, building up on work by Sacchi \cite{sacchi2005} showed that in the above scenario, a ``quantum illumination" transmitter, i.e., one that uses entangled light at the transmitter and an optimal quantum receiver, can perform substantially better than an un-entangled coherent laser transmitter, despite there being no entanglement left between the target-return and the idler beams due to high loss and noise. This is the first example of an entanglement-based performance gain in the bosonic-channel setting where the initial entanglement does not survive the loss and noise in the system. More recently \cite{tan2008}, Tan et. al. showed that using a sequence of identical two-mode-squeezed Gaussian states obtained from spontaneous parametric downconversion (SPDC), in conjunction with an optimal receiver, one may obtain up to a factor of 4 (i.e., 6 dB) gain in the error-probability exponent over an optimum-reception coherent-state radar in a highly lossy and noisy scenario. This optimal receiver can be abstractly expressed as a projective measurement that projects onto the positive eigenspace of the difference of the density operators of the multi-mode states of the target return and retained idler modes, under the two hypotheses -- $H_0$: target absent, and $H_1$: target present. However, no structured receiver design is yet known that can harness any of the 6 dB performance gain. 

We present the design of a structured receiver, which despite being inferior to the (yet un-implemented) optimal joint-detection scheme, in conjunction with the SPDC transmitter is shown to achieve up to a factor of 2 (i.e., 3 dB) error-exponent advantage over the optimum-reception classical sensor in the high loss, low brightness, high noise regime. Our receiver attempts to directly measure the remnant phase-sensitive cross-correlations between the return-idler mode pairs, by mixing the return and the idler beams on a parametric amplifier and photo-detecting the output. Numerical evidence suggests that our receiver achieves the asymptotic error-exponent of the optimal separable measurement, hence indicating that any superior receiver would have to make a complex joint measurement over multiple return-idler mode pairs.

\section{Background}

An SPDC transmitter generates $K$ independent spatio-temporal signal-idler mode pairs $\left\{{\hat a}_S^{(k)}, {\hat a}_I^{(k)}\right\}$; $k \in \left\{1, \ldots, K\right\}$, each prepared in an identical entangled two-mode-squeezed state with a Fock-basis representation
\begin{equation}
|\psi\rangle_{SI} = \sum_{n=0}^{\infty}\sqrt{\frac{N_S^n}{(N_S+1)^{n+1}}}|n\rangle_S|n\rangle_I,
\end{equation}
where $N_S$ is the mean photon number in each signal and idler mode. In the quadrature representation, $|\psi\rangle_{SI}$ is a maximally-entangled zero-mean Gaussian state with second-order quadrature field moments given by
\begin{eqnarray}
\langle{\hat a_{S_m}^{(k)2}}\rangle &=& \langle{\hat a_{I_m}^{(k)2}}\rangle = \frac{2N_S+1}{4}, \text{and} \\
\langle{\hat a_{S_m}^{(k)}}{\hat a_{I_n}^{(k)}}\rangle &=& (-1)^{m+1}\delta_{mn}\frac{\sqrt{N_S(N_S+1)}}{2},
\end{eqnarray}
where ${\hat a_S^{(k)}} = {\hat a_{S_1}^{(k)}} + i{\hat a_{S_2}^{(k)}}$, ${\hat a_I^{(k)}} = {\hat a_{I_1}^{(k)}} + i{\hat a_{I_2}^{(k)}}$, and the standard quadrature field commutators $[{\hat a}_{S_m}^{(k)}, {\hat a}_{I_n}^{(k)}] = 0$ and $[{\hat a}_{S_m}^{(k)}, {\hat a}_{S_n}^{(k)}] = [{\hat a}_{I_m}^{(k)}, {\hat a}_{I_n}^{(k)}] = {(i/2)\delta_{mn}}$ apply for $1 \le k \le K$\footnote{We drop the superscript $(k)$ from the annihilation operators whenever convenient. As each signal-idler mode pair is prepared identically and undergo an identical channel transformation, this will cause no ambiguity in notation.}. Each signal mode is transmitted serially toward a spatial region that may or may not contain a weakly-reflecting specular point target, but in either case, contains a bright thermal-noise bath. Under hypothesis $H_0$ (no target), the target-return mode ${\hat a_R} = {\hat a_B}$, where ${\hat a_B}$ is in a thermal state with mean photon number $N_B \gg 1$. Under hypothesis $H_1$ (target present), ${\hat a_R} = \sqrt{\kappa}{\hat a_S} + \sqrt{1-\kappa}{\hat a_B}$, where the overall channel transmissivity $\kappa \ll 1$, and ${\hat a_B}$ is now in a thermal state with mean photon number $N_B/(1-\kappa)$. Under $H_0$ the joint return-idler state for each of the $K$ mode pairs ${\hat \rho}_{RI}^{(0)}$ is a product of two zero-mean thermal states $({\hat \rho}_{N_B} \otimes {\hat \rho}_{N_S})$ with mean photon numbers $N_B$ and $N_S$ respectively. Under $H_1$, the joint return-idler state for each of the $K$ mode pairs ${\hat \rho}_{RI}^{(1)}$ is a zero-mean Gaussian state with second-order quadrature field moments given by
\begin{eqnarray}
\langle{\hat a_{R_m}^{(k)2}}\rangle &=& \frac{2({\kappa}N_S+N_B)+1}{4}, \label{eq:aR_PS}\\
\langle{\hat a_{I_m}^{(k)2}}\rangle &=& \frac{2N_S+1}{4}, \text{and} \label{eq:aI_PS}\\
\langle{\hat a_{R_m}^{(k)}}{\hat a_{I_n}^{(k)}}\rangle &=& (-1)^{m+1}\delta_{mn}\frac{\sqrt{{\kappa}N_S(N_S+1)}}{2}. \label{eq:aRaI_PS}
\end{eqnarray}

The {\em{binary detection problem}} is to decide between hypotheses $H_0$ and $H_1$ (assuming they are equally likely) with minimum probability of error, using a quantum joint-detection measurement on the state of the $2K$ return-idler mode pairs at the receiver, $({\hat \rho}_{RI}^{(h)})^{(\otimes K)}$; where $h = 0$ or $1$, given hypotheses $H_0$ or $H_1$ respectively.

Helstrom derived the minimum probability of error $P_{e,{\min}}^{(K)}$ for discriminating two quantum states as a function of the number of available copies (or trials) $K$ \cite{helstrom1976},
$
P_{e,{\min}}^{(K)} = \left(1 - \sum_n\gamma_n^{(+)}\right)/2
$, where $\gamma_n^{(+)}$ are the non-negative eigenvalues of $({\hat \rho}_{RI}^{(1)})^{(\otimes K)} - ({\hat \rho}_{RI}^{(0)})^{(\otimes K)}$. Building up on the classical Chernoff bound in classical detection theory \cite{chernoff1952}, recently Audenaert et. al. derived the quantum Chernoff bound (QCB) as an upper bound to $P_{e,{\min}}^{(K)}$, and also showed the QCB to yield the exact asymptotic rate exponent of the minimum error probability \cite{audenaert2007}. Combining that with a (relatively loose) lower bound on $P_{e,{\min}}^{(K)}$ \cite{tan2008}, and defining $Q_s \triangleq {\rm{Tr}}\left(({\hat \rho}_{RI}^{(0)})^s({\hat \rho}_{RI}^{(1)})^{1-s}\right)$ and $Q_{\rm QCB} \triangleq \min_{0 \le s \le 1}Q_s$, we have
\begin{equation}
\frac{1-\sqrt{1-{Q_{0.5}}^{2K}}}{2} \le P_{e,{\min}}^{(K)} \le \frac{1}{2}Q_{\rm QCB}^K \le \frac{1}{2}Q_{0.5}^K,
\label{eq:bounds}
\end{equation}
where the second inequality (QCB) is asymptotically tight as $K \to \infty$. The QCB is customarily represented as $P_{e,{\min}}^{(K)} \le e^{-KR_Q}/2$ in terms of an error-rate exponent $R_Q \triangleq -{\ln}(Q_{\rm QCB})$. The third inequality is a loose upper bound known as the Bhattacharyya bound.

\begin{figure}
\begin{center}
\includegraphics[width=8cm,angle=0]{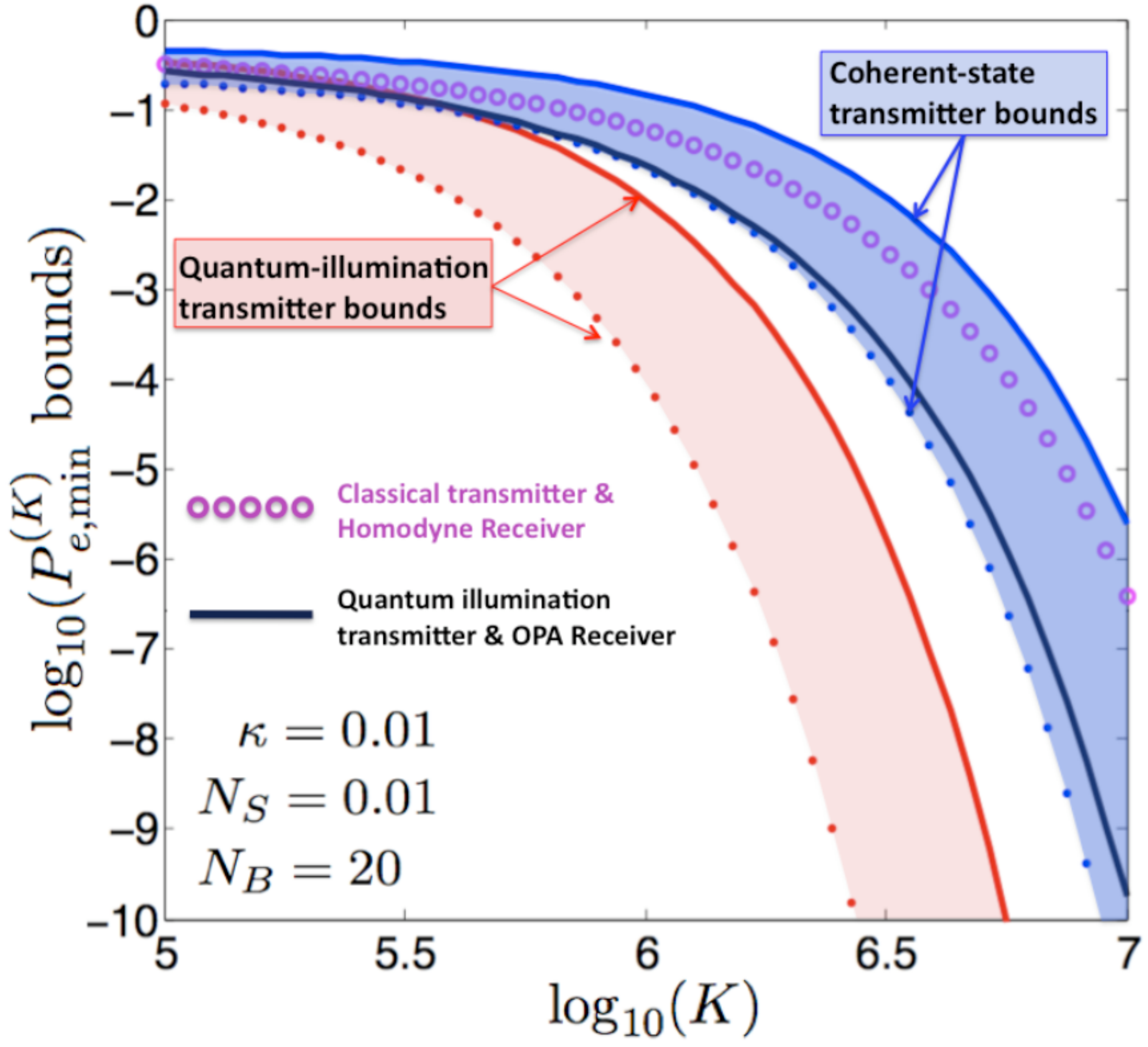}
\end{center}
\caption{Regions between quantum Chernoff upper bounds (solid curves) and (loose) lower bounds (dotted curves) on the minimum error probability \eqref{eq:bounds}, for coherent-state (shaded dark/blue) and quantum-illumination (shaded light/red) transmitters with $K$ transmitted modes each with $N_S = 0.01$ mean transmitted photons per-mode, $N_B = 20$ mean thermal-noise photons per mode, and $\kappa = 0.01$. The coherent-state transmitter lower bound also applies to ALL classical-state transmitters with a total of $KN_S$ photons in all $K$ transmitted modes; thus depicting an undisputed asymptotic advantage obtained by the quantum illumination (SPDC) transmitter with joint optimum reception over all $K$ return-idler mode pairs. The curve plotted with (magenta) circles depicts the error-probability performance of the coherent-state transmitter and mode-by-mode homodyne-detection receiver, which is asymptotically optimal (as $K \to \infty$) for the coherent-state transmitter in the $N_B \gg 1$ regime. The thick black line is the performance of the OPA receiver that we propose in this paper, which performs substantially better compared to the optimal classical sensor (magenta circles) --- note that the magenta circles is the best error-probability performance that can be achieved by {\em any} classical transmitter (the error-probability lower bound of the blue region is rather loose and anything below the magenta circles cannot be achieved by any classical transmitter). The error-probability performance of the (yet unknown) optimum joint-detection receiver for the SPDC transmitter will lie in the (red/light shaded) region between the bounds for the quantum-illumination transmitter, and will hug the quantum-illumination transmitter upper bound for high values of K.}
\label{fig:bounds}
\end{figure}

Symplectic decomposition of Gaussian-state covariance matrices was used to compute the QCB explicitly \cite{pirandola2008}, for both the coherent-state and the entangled (SPDC) transmitters \cite{tan2008}, and it was shown that in the high loss, weak transmission and bright background regime, i.e., with $N_S \ll 1$, $\kappa \ll 1$, and $N_B \gg 1$, the entangled transmitter yields a QCB error-exponent $R_Q = {\kappa}N_S/N_B$, which is four times (or $6$ dB) higher than the error-exponent $R_C = {\kappa}N_S/(4N_B)$ for a coherent-state transmitter with identical per-mode average transmitted photons as the entangled transmitter. In Fig.~\ref{fig:bounds}, we plot the regions captured by the upper and lower bounds in \eqref{eq:bounds} for both the classical and the entangled transmitters, showing a clear advantage of quantum over classical illumination.

\section{Receiver design for classical illumination}

For a coherent-state transmitter, each received mode ${\hat a_R}$ is in a thermal state with mean photon number $N_B$, and a mean-field value $\langle{\hat a_R}\rangle = 0$ or $\sqrt{\kappa{N_S}}$ for hypotheses $H_0$ and $H_1$ respectively. Hence, homodyne detection on each received mode ${\hat a_R^{(k)}}$ yields a variance-$(2N_B+1)/4$ Gaussian-distributed random variable $X_k$ with mean $0$ or $\sqrt{\kappa{N_S}}$ given the hypothesis. Assuming both hypotheses to be equally likely\footnote{With no information about the prior probability of a target being present, we can safely assume the hypotheses to be equally likely. Unequal priors just change the inconsequential pre-factors in actual error-probability and Chernoff bound expressions presented in the paper. Error-exponents remain the same.}, the minimum error probability rule to decide between $H_0$ and $H_1$ is to use the sufficient statistic $X = X_1 + \ldots + X_K$ against a threshold detector, i.e., say $``H_0"$ if $X < (K\sqrt{\kappa{N_S}})/2$ and $``H_1"$ otherwise. The probability of error is given by the Gaussian error-function:
\begin{equation}
P_{e,{\rm{hom}}}^{(K)} = \frac{1}{2}{\rm{erfc}}\left(\sqrt{\frac{\kappa{N_S}K}{4N_B+2}}\right) \approx \frac{1}{2\sqrt{{\pi}KR_{C_{\rm{hom}}}}}e^{-KR_{C_{\rm{hom}}}}, \nonumber
\end{equation}
where the approximation holds for ${\kappa}N_SK/(4N_B+2) \gg 1$ and $R_{C_{\rm{hom}}} ={\kappa{N_S}}/({4N_B+2})$ is the error-exponent. For $N_B \gg 1$, $R_{C_{\rm{hom}}} \approx {\kappa}N_S/4N_B$, identical to the QCB error-exponent for the coherent-state transmitter. It is straightforward to see that for $K$ large enough, $P_{e,{\rm{hom}}}^{(K)} \le e^{-KR_{C_{\rm{hom}}}}/2$. Therefore in the high-background regime, mode-by-mode homodyne detection is asymptotically optimal for the coherent-state transmitter.

\section{Receiver design for quantum illumination}

For the SPDC transmitter, each received return-idler mode pair $\left\{{\hat a}_R, {\hat a}_I\right\}$ is in a joint Gaussian state with zero mean under both hypotheses, and covariance matrix $V = \langle\left[
\begin{array}{cccc}
{\hat a}_R & {\hat a}_I & {\hat a}_R^{\dagger} & {\hat a}_I^\dagger 
\end{array}
\right]^{T}\left[
\begin{array}{cccc}
{\hat a}_R^{\dagger} & {\hat a}_I^\dagger & {\hat a}_R & {\hat a}_I 
\end{array}
\right]\rangle$ whose entries are readily computed using the quadrature field moments in Eqs.~\eqref{eq:aR_PS}, ~\eqref{eq:aI_PS} and ~\eqref{eq:aRaI_PS}. For Hypothesis $H_1$, 

\begin{equation}
V={\tiny{\left[
\begin{array}{cccc}
{\kappa}N_S + N_B + 1 & 0 & 0 & \sqrt{\kappa{N_S(N_S+1)}} \\
0 & N_S+1 & \sqrt{\kappa{N_S(N_S+1)}} & 0 \\
0 & \sqrt{\kappa{N_S(N_S+1)}} & {\kappa}N_S+N_B & 0 \\
\sqrt{\kappa{N_S(N_S+1)}} & 0 & 0 & N_S
\end{array}
\right] \nonumber
}}
\end{equation}
and for hypothesis $H_0$, $V = {\rm{diag}}(N_B+1, N_S+1, N_B, N_S)$. Unlike in the coherent-state transmitter case, the entangled transmitter results in zero-mean joint return-idler states under both hypotheses. As is evident from the covariance matrices under $H_0$ and $H_1$, in the $N_S \ll1$, $\kappa \ll 1$, $N_B \gg 1$ regime, the sole distinguishing factor between the two hypotheses that makes quantum illumination perform superior to the un-entangled coherent-state transmitter, are the off-diagonal terms of $V$ bearing the remnant phase-sensitive cross-correlations of the return-idler mode pairs when the target is present, $\sqrt{\kappa{N_S(N_S+1)}}$. The optimal joint-detection receiver acts on all the $2K$ received return-idler modes and yields the minimum probability of error $P_{e,{\min}}^{(K)}$ by capturing the remnant return-idler phase-sensitive cross-correlations for the $H_1$-hypothesis in the most efficient way. 

\begin{figure}
\begin{center}
\includegraphics[width=9cm,angle=0]{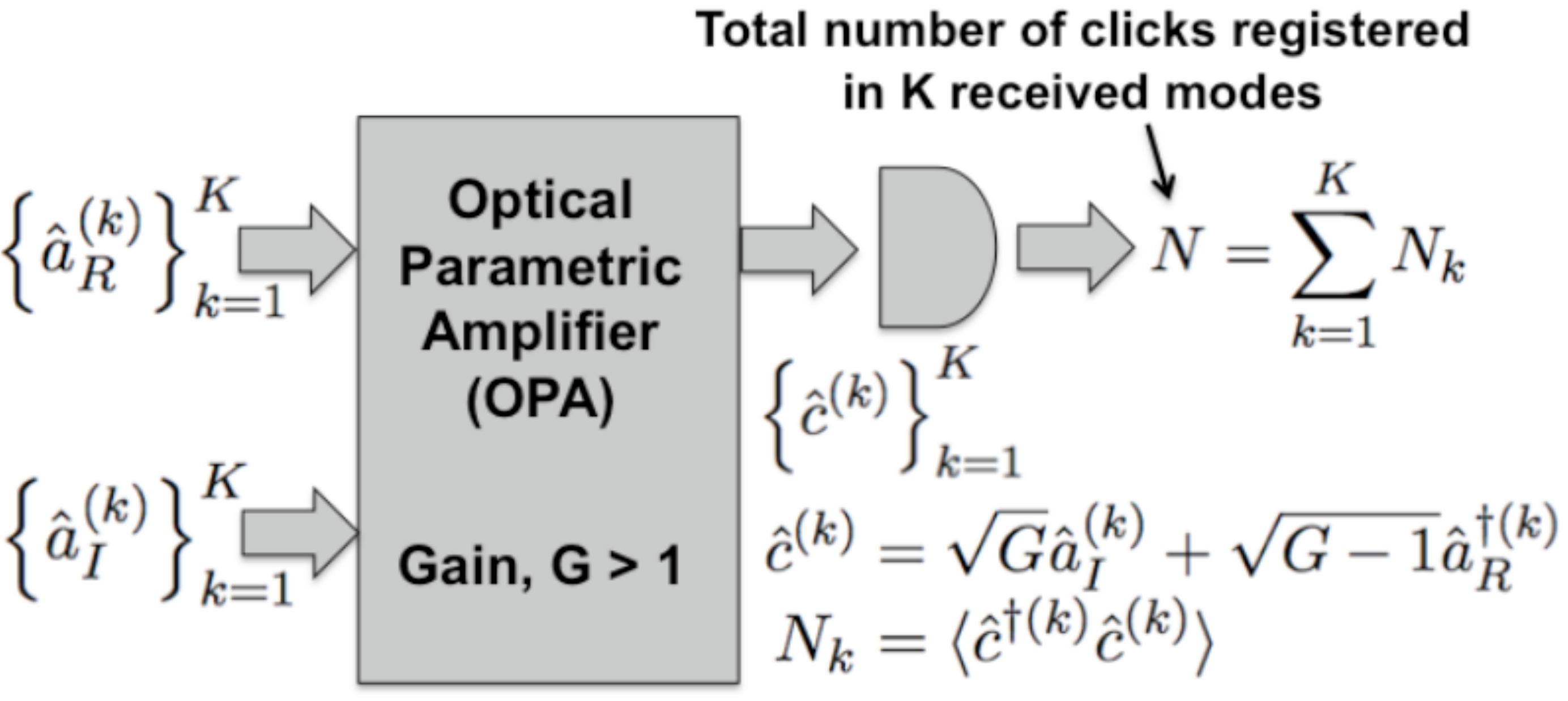}
\end{center}
\caption{A quantum joint-detection receiver that mixes all the received $K$ return-idler mode pairs pairwise on an optical parametric amplifier (OPA) with gain $G$, and counts the total number of clicks $N$ on a photon counter at one output port of the OPA over all $K$ output modes. The receiver decides in favor of hypotheses $H_0$ or $H_1$ depending upon whether $N < N_{\rm{th}}$ or $N \ge N_{\rm{th}}$, $N_{\rm{th}}$ being the decision threshold.} 
\label{fig:OPArcvr}
\end{figure}

The receiver approach we propose uses an optical parametric amplifier (OPA) which is a type-II degenerate amplifier constructed from a $\chi^{(2)}$ non-linear crystal. The incident return and idler modes ${\hat a}_R^{(k)}$ and ${\hat a}_I^{(k)}$ are combined and amplified by an OPA driven by a strong local pump beam, producing pairs of output modes 
\begin{eqnarray}
{\hat c}^{(k)} &=& \sqrt{G}{\hat a}_I^{(k)} + \sqrt{G-1}{\hat a}_R^{\dagger{(k)}} \text{ and} \\
{\hat d}^{(k)} &=& \sqrt{G}{\hat a}_R^{(k)} + \sqrt{G-1}{\hat a}_I^{\dagger{(k)}},
\end{eqnarray} 
where $G > 1$ is the gain of the OPA (see Fig.~\ref{fig:OPArcvr}). Each output mode ${\hat c}$ is in a zero-mean thermal state with mean photon number under the two hypotheses given by
\begin{equation}
\langle{\hat c}^\dagger{\hat c}\rangle = \left\{
\begin{array}{c} GN_S + (G-1)(1+N_B) \triangleq N_0, \quad H_0 \\ \\
GN_S + (G-1)(1+N_B + {\kappa}N_S) \\ 
+ 2\sqrt{G(G-1)}\sqrt{{\kappa}N_S(N_S+1)} \triangleq N_1, \quad H_1 \end{array}\nonumber
\right.
\end{equation}
i.e., ${\hat \rho}_c = \sum_{n=0}^{\infty}(N_m^n/(1+N_m)^{1+n})|n\rangle\langle{n}|$, for $m \in \left\{0, 1\right\}$ for $H_0$ and $H_1$ respectively. Hence, the joint state of the $K$ received modes ${\hat c}^{(k)}$ is a $K$-fold tensor product ${\hat \rho}_c^{\otimes{K}}$ of identical zero-mean thermal states with per-mode mean photon number $N_0$ or $N_1$ depending upon which of the two hypothesis is true. A $K$-fold product of thermal states is diagonal in the $K$-fold tensor-product of photon-number bases of the $K$ modes. Hence the optimum joint quantum measurement to distinguish between the two hypotheses is to count photons on each output mode ${\hat c}^{(k)}$ and decide between the two hypotheses based on the total photon count $N$ over all $K$ detected modes, using a threshold detector. The probability mass function of $N$ under the two hypotheses is given by
\begin{equation}
P_{N|H_m}(n|H_m) = \frac{1}{(1+N_m)^K}\left(\begin{array}{c}n+K-1\\n\end{array}\right)\left(\frac{N_m}{1+N_m}\right)^n, \nonumber
\end{equation}
where $m = 0$ or $1$ based on which hypothesis is true. The mean and variance of this distribution are $KN_m$ and $K\sigma_m^2$ respectively, where $\sigma_m^2 = N_m(N_m+1)$. For large $K$ the above conditional distributions for $N$ approach Gaussian distributions (due to the central limit theorem (CLT)) with means and variances given by $KN_m$ and $K\sigma_m^2$ respectively. The probability of error is given by
\begin{equation}
P_{e,{\rm{OPA}}}^{(K)} = \frac{1}{2}{\rm{erfc}}\left(\sqrt{R_{{\rm{OPA}}}K}\right) \approx \frac{1}{2\sqrt{{\pi}KR_{\rm{OPA}}}}e^{-KR_{{\rm{OPA}}}}, \nonumber
\end{equation}
where an error-exponent $R_{{\rm{OPA}}} = (N_1-N_0)^2/2(\sigma_0+\sigma_1)^2$ can be achieved using a threshold detector that decides in favor of hypotheses $H_0$ or $H_1$ depending upon whether $N < N_{\rm{th}}$ or $N \ge N_{\rm{th}}$, with $N_{\rm{th}} = \lceil{K(\sigma_1N_0+\sigma_0N_1)/(\sigma_0+\sigma_1)}\rceil$. $R_{{\rm{OPA}}}$ is a function of the OPA gain $G$. Given $N_S \ll 1$, $\kappa \ll 1$, $N_B \gg 1$, intuitively a small gain, $G = 1 + \epsilon^2$ with $0 < \epsilon \ll 1$ will be optimal to distinguish between the two hypotheses, such that the difference in the total mean photon count is dominated by the term proportional to the phase-sensitive correlation, i.e., $K(N_1-N_0) = K({\epsilon^2}{\kappa}N_S + 2{\epsilon}\sqrt{1+\epsilon^2}\sqrt{\kappa{N_S(N_S+1)}}) \approx 2{\epsilon}K\sqrt{\kappa{N_S(N_S+1)}}$, for $N_S \ll 1$, $\kappa \ll 1$, $N_B \gg 1$ and $\epsilon \ll 1$. For the problem parameters as chosen in Fig.~\ref{fig:bounds}, i.e., $N_S = 0.01$, $\kappa = 0.01$ and $N_B = 20$, $R_{{\rm{OPA}}}$ is maximized for $G = 1+ 5 \times 10^{-3}$, confirming our intuition that a small gain is optimal. For this gain, $R_{{\rm{OPA}}}  = 2 \times 10^{-6}$. For the above parameters, error-exponents for the classical and SPDC transmitters with optimal measurement are $R_C = 1.25 \times 10^{-6}$ and $R_Q = 5 \times 10^{-6}$ respectively. Hence for these parameters, our receiver gets $\approx 2$ dB gain in the error-exponent over the classical system. 

Finally, to establish the asymptotic error-exponent performance of the OPA receiver, we will use the classical Bhattacharyya bound to the error-probability $P_{e,{\rm{OPA}}}^{(K)}$. The Bhattacharyya bound to the error probability in distinguishing between the two distributions $P_{N|H_0}(n|H_0)$ and $P_{N|H_1}(n|H_1)$ using K i.i.d. observations is given by:

\begin{equation}
P_{e,{\rm{OPA}}}^{(K)} \le \frac{1}{2}Q_B^K,
\end{equation} 
where 
\begin{eqnarray}
Q_B &=& \sum_{n=0}^{\infty}\sqrt{p_{N|H_0}(n|H_0)p_{N|H_1}(n|H_1)} \\
&=& \frac{1}{\sqrt{(1+N_0)(1+N_1)}-\sqrt{N_0N_1}}.
\end{eqnarray} 
Using the approximation $N_1-N_0 = 2{\epsilon}\sqrt{\kappa{N_S(N_S+1)}} \triangleq \delta \ll 1$, we obtain 
\begin{eqnarray}
Q_B &=& \frac{1}{\sqrt{(1+N_0)(1+N_1)}-\sqrt{N_0N_1}} \\
&=& \frac{1}{(1+N_0)\sqrt{1+\frac{\delta}{1+N_0}} - N_0\sqrt{1+\frac{\delta}{N_0}}} \\
&=& \left[ (1+N_0)\left(1+ \frac{\delta}{2(1+N_0)} + \frac{\delta^2}{8(1+N_0)^2}\right) \right. \nonumber\\
&& \left.- N_0\left(1+\frac{\delta}{2N_0} + \frac{\delta^2}{8N_0}\right)  \right]^{-1} \\
&=& \frac{1}{1 + \frac{\delta^2}{8N_0(1+N_0)}} \\
&\approx& 1 - \frac{(N_1-N_0)^2}{8N_0(1+N_0)} \\
&=& 1 - R_B,
\end{eqnarray} 
with $R_B = (N_1-N_0)^2/(8N_0(1+N_0))$. This translates to the bound 
\begin{equation}
P_{e,{\rm{OPA}}}^{(K)} \le \frac{1}{2}e^{-KR_B}, 
\end{equation}
where the Bhattacharyya bound exponent $R_B$ is given by 
\begin{eqnarray}
R_B &=& \frac{{\epsilon}^2\kappa{N_S(N_S+1)}}{2N_S(N_S+1)+2{\epsilon^2}(1+2N_S)(1+N_S+N_B)} \nonumber \\
&\approx& \kappa{N_S}/2N_B,
\end{eqnarray}
for a choice of $\epsilon^2 = N_S/\sqrt{N_B}$, for $N_S \ll 1$, $\kappa \ll 1$, $N_B \gg 1$ ($\epsilon \ll 1$ is automatically satisfied)\footnote{A different $\epsilon$ satisfying $N_S/N_B \ll \epsilon^2 \ll 1/N_B$ would work as well.}. As $R_{\rm{OPA}} \ge R_B$, therefore by construction, for a weak transmitter operating in a highly lossy and noisy regime, the OPA receiver achieves at least a $3$ dB gain in error exponent over the optimum-receiver classical sensor whose QCB error exponent $R_C = \kappa{N_S}/4N_B$. For $N_S \ll 1$ and $\epsilon \ll 1$, both $N_0$ and $N_1 \ll 1$. Hence, a single-photon detector (as opposed to a full photon-counting measurement) suffices to achieve the performance of the receiver depicted in Fig.~\ref{fig:OPArcvr}.

\begin{figure}
\begin{center}
\includegraphics[width=9cm,angle=0]{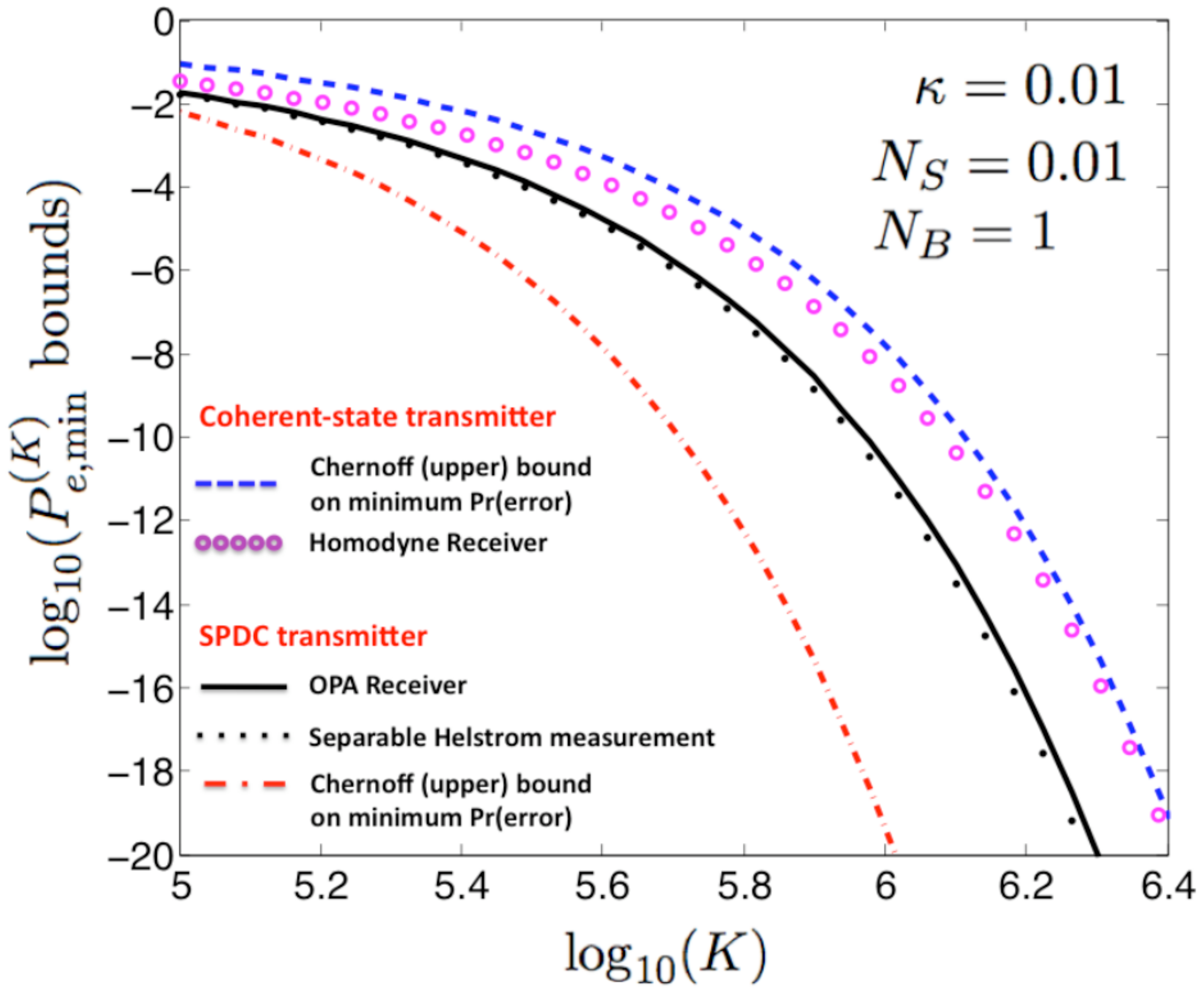}
\end{center}
\caption{Symbol-by-symbol (separable) Helstrom minimum error-probability measurement on each return-idler mode pair seems to have identical error-exponent as (and performs only slightly better than) the OPA receiver. The separable Helstrom measurement performance was calculated by explicitly evaluating the single-shot minimum error-probability $P_{e,{\min}}^{(1)} = \left(1 - \sum_n\gamma_n^{(+)}\right)/2$, where $\gamma_n^{(+)}$ are the non-negative eigenvalues of ${\hat \rho}_{RI}^{(1)} - {\hat \rho}_{RI}^{(0)}$, using CLT to compute the error-probability with independent measurements on all return-idler mode pairs, followed by majority-vote hard decision.}
\label{fig:Helstrom_compare}
\end{figure}

\section{Conclusions}

Entanglement has proven to be one of the most useful resources in quantum information, with applications to improving classical communication rates over noisy quantum channels, teleportation of unknown quantum states over long distances, and in quantum algorithms that can potentially solve certain problems (such as factoring) faster than the corresponding best known conventional classical algorithms. Whereas realizing useful quantum computing is not practically feasible with current technology, exploiting quantum effects to build optical communications and sensing systems that could perform better than corresponding classical systems seems to be well within the reach of current state-of-the-art in experimental quantum optics.

``Quantum illumination" is a novel preliminary attempt at using an entangled source of light to detect the presence of a weakly-reflecting target in the far-field subject to a very lossy and noisy channel \cite{lloyd2008}. Even though the efficacy of using spontaneous parametric downconversion -- a well-known entanglement source -- as a transmitter to obtain a significant advantage (6 dB in error-exponent) over using a classical laser transmitter for target detection was established by means of quantum Chernoff bounds \cite{tan2008}, a structured receiver design that may harness this advantage is not known to date. Our work puts forth an explicit receiver design using conventional quantum optics, that when used along with an SPDC transmitter is capable of obtaining up to a 3 dB error-exponent gain over the optimum-reception classical sensor.

There are several problems and challenges -- both in theory and experiments -- that remain open. It is still not clear whether a sequence of identically prepared two-mode-squeezed states constitutes the best quantum-illumination transmitter. It will hence be of interest to explore whether a transmitter that uses non-Gaussian entangled states, or complex higher-order entanglement between all the transmitted signal-idler mode pairs can achieve more than just a constant-factor improvement in the error exponent. It would also be interesting to extend this work to detecting and imaging spatially extended speckled targets. 

In terms of receiver design, the problem of constructing a receiver for the SPDC transmitter that achieves the full 6 dB error-exponent gain is certainly at the forefront. Numerical evidence suggests that any alternative receiver that makes independent measurements on each return-idler mode pair is not likely to perform significantly better than the OPA receiver in the absence of classical feedforward or soft-decision post-processing through the $K$ measurement instances\footnote{The author thanks Baris I. Erkmen, JPL for pointing out that by phase-conjugating each received mode and detecting the conjugated return and retained idler modes on a balanced dual detector would achieve the same factor-of-$2$ improvement in the error-exponent as the OPA receiver.}, thus necessitating a superior receiver to make a joint measurement over {\em{all}} return-idler mode pairs (see Fig.~\ref{fig:Helstrom_compare})\footnote{To see the details of the calculations, see Appendix A}. The current model assumes the receiver to have complete knowledge of the signal power, channel loss and noise power for the optimal design of the receiver. Whereas the knowledge of the transmitted signal power at the receiver is a reasonable assumption, a full estimation-theoretic study of the performance of quantum illumination with no prior knowledge of channel loss and noise, would be necessary in order to design a realistic prototype of the quantum sensor. 

\section*{Acknowledgments}
The author thanks Jeffrey H. Shapiro, MIT for valuable discussions and for pointing out an error in the second-order moment calculations. The author also thanks Zachary Dutton, BBN, Seth Lloyd, MIT and Baris I. Erkmen, JPL for several interesting discussions on this topic and for feedback on this draft. The author thanks the DARPA Quantum Sensors Program and BBN Technologies Corporation for supporting this research.

\section*{Appendix A}

In order to substantiate our claim about the performance of single-shot Helstrom measurement (depicted in Fig.~\ref{fig:Helstrom_compare}), we compute the error-probability performance of the pair-wise (separable) hard-decision Helstrom measurement (the quantum measurement that minimizes the single-shot probability of error) followed by a majority-vote detector after independently detecting all $K$ return-idler mode pairs. In order to compute the Helstrom measurement minimum probability of error $P_{e,{\rm{min}}}^{(1)}$, we need to compute the non-negative eigenvalues of the difference of the density operators of the two-mode return-idler states under the two hypotheses, i.e. ${\hat \rho}_{RI}^{(1)} - {\hat \rho}_{RI}^{(0)}$. For our propagation and noise model, these two return-idler density operators can be computed in full generality in terms of their respective tensor-product Fock-state basis elements. For a pure-state two-mode entangled transmitter of the form
\begin{equation}
|\psi\rangle_{SI} = \sum_{n=0}^{\infty}\sqrt{p_n}|n\rangle_S|n\rangle_I,
\end{equation}
the return-idler state matrix-elements come out to be:
\begin{equation} 
{}_I\langle{m_2}|_R\langle{m_1}|{\hat \rho}_{RI}^{(0)}|n_1\rangle_R|n_2\rangle_I = \frac{N_B^{n_1}}{(1+N_B)^{n_1+1}}p_{n_2}\delta_{m_1n_1}\delta_{m_2n_2} \nonumber,
\end{equation}
and
\begin{eqnarray} 
{}_I\langle{m_2}|_R\langle{m_1}|{\hat \rho}_{RI}^{(1)}|n_1\rangle_R|n_2\rangle_I = \sqrt{\frac{n_1!n_2!}{(n_1+l)!(n_2+l)!}} \nonumber \\
\times \sqrt{p_{n_2+l}p_{n_2}}\kappa^{l/2}\frac{(n_1+n_2+l)!}{n_1!n_2!}\frac{(N_B+1-\kappa)^{n_2}N_B^{n_1}}{(N_B+1)^{n_1+n_2+l+1}} \nonumber \\
\times {}_2F_1\left[-n_1, -n_2, -(n_1+n_2+l), 1-\frac{\kappa}{N_B(N_B+1-\kappa)}\right], \nonumber
\end{eqnarray}
when $l = m_1-n_1=m_2-n_2$ is satisfied and ${}_I\langle{m_2}|_R\langle{m_1}|{\hat \rho}_{RI}^{(1)}|n_1\rangle_R|n_2\rangle_I = 0$ for $m_1-n_1 \ne m_2-n_2$. $|n\rangle_R$ and  $|n\rangle_I$ for $n \in \left\{0, 1, \ldots, \infty\right\}$ are the photon-number states for the return and the idler mode respectively, and each form a complete orthonormal set of bases for the respective state spaces. ${}_2F_1$ is the well-known hypergeometric function. Using the above expressions for the density operators, we computed the error-probability performance of the symbol-by-symbol (separable) Helstrom minimum error-probability measurement on each return-idler mode pair, and numerical results seem to suggest that this measurement has the same error-exponent as (and performs only slightly better than) the OPA receiver. The separable Helstrom measurement performance was calculated by evaluating the single-shot minimum error-probability $P_{e,{\min}}^{(1)} = \left(1 - \sum_n\gamma_n^{(+)}\right)/2$, where $\gamma_n^{(+)}$ are the non-negative eigenvalues of ${\hat \rho}_{RI}^{(1)} - {\hat \rho}_{RI}^{(0)}$, and then using the central limit theorem to compute the error-probability with independent measurements on all return-idler mode pairs, followed by a majority-vote hard decision (see Fig.~\ref{fig:Helstrom_compare}).

\end{document}